# Ensuring Disturbance Rejection Performance by Synthesizing Grid-Following and Grid-Forming Inverters in Power Systems

Fuyilong Ma, Huanhai Xin*, Zhiyi Li, Linbin Huang

*Abstract*—To satisfy dynamic requirements of power systems, it is imperative for grid-tied inverters to ensure good disturbance rejection performance (DRP) under variable grid conditions. This letter discovers and theoretically proves that for general networks, synthesizing grid-following (GFL) inverters and grid-forming (GFM) inverters can always more effectively ensure the DRP of multiple inverters, as compared to homogeneous inverter-based systems that solely utilize either GFL or GFM inverters. The synthesis of GFL inverters and GFM inverters can concurrently increase the smallest eigenvalue and decrease the largest eigenvalue of the network grounded Laplacian matrix. This can be equivalent to rematching the proper short-circuit ratio (SCR) for GFL and GFM inverters, thereby ensuring the DRP of inverters both in weak and strong grids. The results reveal the necessity of synthesizing diverse inverter control schemes from the network-based perspective. Sensitivity function-based tests and real-time simulations validate our results.

## I. INTRODUCTION

In the pursuit of decarbonizing electrical energy, the large-scale integration of power electronic inverters into the electrical grid becomes essential, serving as the interfaces for the renewable energy sources. Commonly, grid-following (GFL) inverter and grid-forming (GFM) inverter are two typical control schemes for the inverters [1].

Given the fact that external disturbances, including a step change in the load or harmonic disturbances, are inevitable from the public grid [2], it is imperative for the inverter controllers to exhibit disturbance rejection performance (DRP) [3]. This is essential to satisfy the dynamic requirements of power systems [4]. For example, a GFL inverter is expected to maintain the high-quality terminal voltage waveform for the power system, even in the event of disturbance [4]. However, achieving this becomes a challenge in a weak grid (i.e., a grid with the low short-circuit ratio (SCR)), and it could lead to oscillation instability issues [4],[5]. In comparison to the GFL inverter, the GFM inverter has attracted increasing research interest due to its ability to support the grid frequency and voltage, as well as their superior performance in weak grids. However, under strong grid conditions, the GFM control performance may be diminished [6].

Both the existing GFL and GFM controllers generally consider a single inverter-based system (SIBS), but provides no assurance on the DRP of multiple inverters. It has been revealed that the interactions among GFL inverters could deteriorate the overall performance in response to the disturbance, compared to the SIBS [7],[8]. Also, the similar issues have been found in homogeneous systems of GFM inverters [9]. It is challenging to ensure the DRP of multiple GFL or GFM inverters under variable grid conditions. Some studies apply a combination of GFL and GFM controllers to tackle this concern, the effectiveness of which has been verified in a simple two-bus system [10]. However, it is still necessary to give a theoretical explanation for the more general network topology.

To fill this gap, this letter provides a network-based insight, demonstrating that the DRP of the hybrid GFL-GFM inverter system can always be at least superior to that of a homogeneous system composed exclusively of either GFL or GFM inverters, for general networks. Instead of numerical tests, we will present an analytical proof and further reveal the necessity of synthesizing diverse inverter control schemes to ensure the DRP of multiple inverters under variable grid conditions. Frequency- and time-domain tests from an experimental system are presented to validate the analysis.

## II. PROBLEM FORMULATION

Without the loss of generality, we examine a set of $n$ inverter nodes, indexed by $i \in \{1,...,n\}$, dynamically coupled though an ac network. Consider three multi-inverter systems, denoted as $\Gamma_1$, $\Gamma_2$ and $\Gamma_3$. Each system shares the same network, but with different types of inverters (i.e., GFL or GFM inverters) connected to their nodes. Specifically, $\Gamma_1$ and $\Gamma_2$ represent homogeneous systems of GFL and GFM inverters, respectively, while $\Gamma_3$ represent a hybrid system with identical GFL inverters (connected to node $1 \sim n_1$) and identical GFM inverters (connected to node $n_1+1 \sim n_2$), $n_1 + n_2 = n$. The linearized closed-loop dynamics of the three systems can be represented by the block diagram in Fig.1 [7].

As depicted in Fig.1, $Y_N(s) = \boldsymbol{B} \otimes F(s)$ represents the admittance model for the network dynamics [7], which is same in all three systems; $\boldsymbol{B} \in \mathbb{R}^{n \times n}$ represents the network susceptance matrix [7] and also is the network grounded Laplacian matrix that preserves $n$ inverter nodes and eliminates infinite nodes and interior nodes [11]; $F(s) = [s, \omega_0; -\omega_0, s]/(s^2/\omega_0 + \omega_0)$ ; $Y_G(s)$ represents the block-diagonal transfer function of admittance models of inverters, that is,

$$\boldsymbol{Y}_G^{\Gamma_1}(s) = \mathbb{I}_n \otimes Y_{GFL}(s), \boldsymbol{Y}_G^{\Gamma_2}(s) = \mathbb{I}_n \otimes Y_{GFM}(s),$$
$$\boldsymbol{Y}_G^{\Gamma_3}(s) = \begin{bmatrix} \mathbb{I}_{n_1} \otimes Y_{GFL}(s) & \\ & \mathbb{I}_{n_2} \otimes Y_{GFM}(s) \end{bmatrix}, \quad (1)$$

where $\mathbb{I}_n$ represents a $n$-by-$n$ identity matrix; $\otimes$ denotes Kronecker product; $Y_{GFL}(s)$ and $Y_{GFM}(s)$ are the 2×2 admittance transfer function matrices of the GFL and GFM inverter, respectively, whose detailed expressions can be found in [12]; $\Delta \boldsymbol{I}_{per}$ and $\Delta \boldsymbol{U}_{per}$ represent the external voltage disturbance in series with each branch and current disturbance in parallel with each node for inverters, respectively, which can be equivalent to the disturbances from the public grids in terms of small disturbance; $\Delta \boldsymbol{I}$ and $\Delta \boldsymbol{U}$ represent the current and voltage response vectors of inverters following the disturbance, respectively.

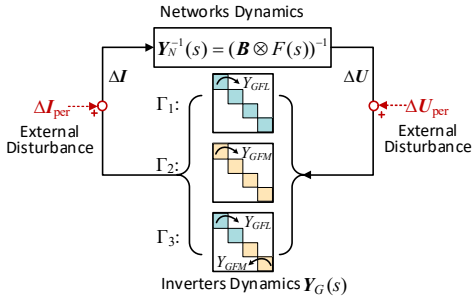

Fig.1 Block diagram of closed-loop systems with multiple GFL or GFM inverters under the external disturbance.

The multi-inverter systems in Fig.1 are the multivariable feedback systems [3]. In the presence of the external disturbances denoted as $\Delta \boldsymbol{U}_{per}(s)$ or $\Delta \boldsymbol{I}_{per}(s)$ in Fig.1, the DRP of the systems can be quantified by using the sensitivity function, as it mirrors the gain of the closed-loop system in response to the disturbance [3]. Let $s = j\omega$, and the sensitivity function can be formulated as

$$\boldsymbol{S}(j\omega) = (\mathbb{I}_{2n} + \boldsymbol{L}(j\omega))^{-1}, \boldsymbol{L}(j\omega) = \boldsymbol{Y}_N^{-1}(j\omega)\boldsymbol{Y}_G(j\omega), \quad (2)$$

$$\forall \omega : \bar{\sigma}\{\boldsymbol{S}(j\omega)\} = 1/\underline{\sigma}\{\mathbb{I}_{2n} + \boldsymbol{L}(j\omega)\}, \quad (3)$$

$$\kappa_P = \max_\omega \bar{\sigma}(\boldsymbol{S}(j\omega)) = \max_\omega \{1/\underline{\sigma}\{\mathbb{I}_{2n} + \boldsymbol{L}(j\omega)\}\} = \|\boldsymbol{S}(j\omega)\|_\infty, \quad (4)$$

where $\boldsymbol{S}(j\omega)$ and $\boldsymbol{L}(j\omega)$ represent the sensitivity function and the open-loop transfer function of systems in Fig.1, respectively; $\bar{\sigma}\{\cdot\}$ and $\underline{\sigma}\{\cdot\}$ denote the maximum and minimum singular values of a matrix, respectively; $\bar{\sigma}\{\boldsymbol{S}(j\omega)\}$ can quantify the DRP of the system under the external disturbance at the frequency point $\omega$; $\kappa_P$ represents the sensitivity peak and can be the DRP indicator within the wide frequency range; the lower $\kappa_P$ means the better DRP of the system [3]; $\|\cdot\|_\infty$ denotes $\mathcal{H}_\infty$ norm.

Based on the above formulations, we now articulate the interested problem as follows.

**Problem.** Could the synthesis of GFL and GFM inverters in system $\Gamma_3$ more effectively ensure the DRP of multiple inverters, compared to homogeneous inverter-based systems like system $\Gamma_1$ or system $\Gamma_2$, which solely utilize either GFL or GFM inverters? That is, do we have $\kappa_P^{\Gamma_3} < \max_\omega \{\kappa_P^{\Gamma_1}, \kappa_P^{\Gamma_2}\}$ ?

## III. MAIN RESULTS

### A. Relationship Between DRP and SCR

Considering that the inverter control performance is commonly related to the grid conditions [1], we firstly illustrate the role of SCR in the DRP of a GFL-based or GFM-based SIBS. The sensitivity function and its peak of SIBS can be formulated as

$$\boldsymbol{S}^{SIBS}(j\omega) = (\mathbb{I}_2 + \boldsymbol{L}^{SIBS}(j\omega))^{-1}, \boldsymbol{L}^{SIBS}(j\omega) = \lambda_{SCR}^{-1} F^{-1} Y_{GFL/GFM}(j\omega) \quad (5)$$

$$\kappa_P^{SIBS} = \max_\omega \{1/\underline{\sigma}\{\mathbb{I}_2 + \lambda_{SCR}^{-1} F^{-1}(j\omega) Y_{GFL/GFM}(j\omega)\}\} \quad (6)$$

where $\boldsymbol{S}^{SIBS}(j\omega)$ and $\boldsymbol{L}^{SIBS}(j\omega)$ represent the sensitivity function and the open-loop transfer function of SIBS, respectively; $\kappa_P^{SIBS}$ represents the sensitivity peak of SIBS; $\lambda_{SCR} = L_{line}^{-1}$ represents the SCR [7]. $L_{line}$ represents the grid inductance in the SIBS.

Fig. 2 shows curves of the maximum singular value of sensitivity function $\bar{\sigma}\{\boldsymbol{S}^{SIBS}(j\omega)\}$ for the GFL/GFM-based SIBSs in (5) with SCR varying from 3.0 to 7.0. Parameters of inverters can be founded in TABLE A.I in the Appendix A. It can be seen from Fig. 2 that $\kappa_P^{SIBS}$ in GFL-based SIBS varies from 29.1 dB to 6.2dB and $\kappa_P^{SIBS}$ in GFM-based SIBS varies from 5.1dB to 10.7dB with SCR increasing. Thus, the DRP of GFL-based SIBS is improved with SCR increasing, while the DRP of GFM-based SIBS is deteriorated, which are consistent with empirical views [1],[6].

As a result, SCR can be served as a network-based indicator for quantifying the DRP of the SIBS based on a GFL or GFM inverter, comparable to the sensitivity peak.

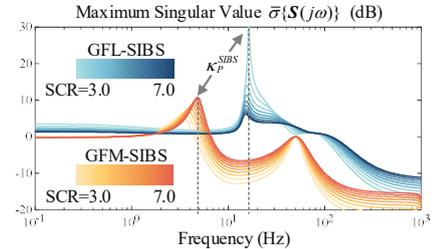

Fig. 2 Curves of the maximum singular value of sensitivity function for GFL-based and GFM-based SIBSs with SCR increasing from 3.0 to 7.0.

### B. DRP Quantification of Homogeneous Systems

Based on the preceding analysis in the SIBS, this subsection illustrates that the DRP of homogeneous systems of GFL or GFM inverters (i.e., systems $\Gamma_1$ and $\Gamma_2$) can be quantified by the SCR in the equivalent SIBSs of these two homogeneous systems.

The maximum singular value of sensitivity functions of systems $\Gamma_1$ or $\Gamma_2$ in (3) can be rewritten as

$$\bar{\sigma}\{\boldsymbol{S}(j\omega)\} = 1/\underline{\sigma}\{\mathbb{I}_{2n} + \boldsymbol{Y}_N^{-1}(j\omega)\boldsymbol{Y}_G^{\Gamma_1/\Gamma_2}(j\omega)\}$$
$$= 1/\underline{\sigma}\{\mathbb{I}_{2n} + \boldsymbol{B}^{-1} \otimes F^{-1} Y_{GFL/GFM}(j\omega)\} \quad (7)$$

Notice that the network grounded Laplacian matrix $\boldsymbol{B}$ in (7) is a positive definite matrix, and thus existing a unity matrix $\boldsymbol{W} \in \mathbb{R}^{n \times n}$ makes

$$\boldsymbol{W}^* \boldsymbol{B} \boldsymbol{W} = \boldsymbol{\Lambda} = diag\{\lambda_1, ..., \lambda_n\}, \boldsymbol{W}^* \boldsymbol{W} = \mathbb{I}_n \quad (8)$$

where $\boldsymbol{\Lambda}$ is a diagonal matrix whose elements satisfy $0 < \lambda_1 \leq ... \leq \lambda_i \leq ... \leq \lambda_n$, $i \in \{1,...,n\}$ ; $\lambda_1 = \underline{\lambda}\{\boldsymbol{B}\}$ and

$\lambda_n = \bar{\lambda}\{\boldsymbol{B}\}$ represent the smallest and largest eigenvalues of the network grounded Laplacian matrix $\boldsymbol{B}$, respectively; $(\cdot)^*$ represents the conjugate transpose.

Since singular values are invariant for a unity transform [14], the maximum singular value of the sensitivity function in (7) can be rewritten as (for simplicity, the following derivations omit $j\omega$)

$$\begin{aligned}\bar{\sigma}\{\boldsymbol{S}\} &= 1/\underline{\sigma}\{(\boldsymbol{W}^* \otimes \mathbb{I}_2)(\mathbb{I}_{2n} + \boldsymbol{B}^{-1} \otimes F^{-1}Y_{GFL/GFM})(\boldsymbol{W} \otimes \mathbb{I}_2)\}\\ &= 1/\underline{\sigma}\{\mathbb{I}_{2n} + \boldsymbol{\Lambda}^{-1} \otimes F^{-1}Y_{GFL/GFM}\}\\ &= \max\{1/\underline{\sigma}\{\mathbb{I}_2 + \lambda_1^{-1} F^{-1}Y_{GFL/GFM}\},...\\ &\quad ,1/\underline{\sigma}\{\mathbb{I}_2 + \lambda_n^{-1} F^{-1}Y_{GFL/GFM}\}\}\end{aligned} \quad (9)$$

By combing (6) and (9), the DRP of systems $\Gamma_1$ and $\Gamma_2$ can be quantified by the eigenvalues $\lambda_i, i \in \{1,...,n\}$ which can be regarded as SCR of $n$ equivalent SIBSs in (9). It is noteworthy that $\lambda_1 = \underline{\lambda}\{\boldsymbol{B}\}$ can be regarded as the so-called generalized SCR (gSCR) when the inverter capacities are rated [7].

More specifically, by combing the illustration from Section III.A, the DRP of system $\Gamma_1$ with GFL inverters can be determined by that of SIBS with the lowest SCR (i.e., $\lambda_1 = \underline{\lambda}\{\boldsymbol{B}\}$), and the DRP of system $\Gamma_2$ with GFM inverters can be determined by that of SIBS with the highest SCR (i.e., $\lambda_n = \bar{\lambda}\{\boldsymbol{B}\}$). Thus, we have

$$\begin{aligned}\kappa_P^{\Gamma_1} &= \max_\omega \{1/\underline{\sigma}\{\mathbb{I}_2 + \lambda_1^{-1} F^{-1} Y_{GFL}\}\},\\ \kappa_P^{\Gamma_2} &= \max_\omega \{1/\underline{\sigma}\{\mathbb{I}_2 + \lambda_n^{-1} F^{-1} Y_{GFM}\}\}.\end{aligned} \quad (10)$$

### C. Comparison with Synthesizing GFL and GFM Inverters

This subsection quantifies and compares the DRP of the hybrid GFL-GFM inverter system (i.e., system $\Gamma_3$) with that of the homogeneous systems composed solely of GFL and GFM inverters (i.e., system $\Gamma_1$ and $\Gamma_2$).

Considering the interactions between GFL and GFM inverters, we firstly demonstrate how the hybrid GFL-GFM inverter system can be decoupled into two homogeneous subsystems of GFL and GFM inverters. According to [13], the closed-loop characteristic equation of system $\Gamma_3$ in Fig.1 can be formulated based on that of two homogeneous subsystems 1 and 2 as

$$0 = \det\{\mathbb{I}_{2n} + \boldsymbol{L}^{\Gamma_3}\} = \det\{\mathbb{I}_{2n_1} + \boldsymbol{L}^{sub1}\}\det\{\mathbb{I}_{2n_2} + \boldsymbol{L}^{sub2}\} \quad (11)$$

where

$$\begin{aligned}\boldsymbol{L}^{\Gamma_3} &= \boldsymbol{Y}_G^{\Gamma_3} \boldsymbol{Y}_N^{-1}, \boldsymbol{L}^{sub1} = (\mathbb{I}_{n_1} \otimes Y_{GFL})(\boldsymbol{Y}_N^{sub1})^{-1},\\ \boldsymbol{L}^{sub2} &= (\mathbb{I}_{n_2} \otimes Y_{GFM})(\boldsymbol{Y}_N^{sub2})^{-1},\end{aligned} \quad (12)$$

where $\boldsymbol{L}^{sub1}$ and $\boldsymbol{L}^{sub2}$ represent the open-loop transfer function matrices of the homogeneous subsystems 1 and 2 of system $\Gamma_3$, respectively; $\boldsymbol{Y}_N^{sub1}$ and $\boldsymbol{Y}_N^{sub2}$ represent the network admittance transfer function matrices of homogeneous subsystems 1 and 2, respectively, as follows

$$\boldsymbol{Y}_N^{sub1} = \boldsymbol{B}[n_1, n_1] \otimes F - (\boldsymbol{B}[n_1, n_2] \otimes F)(\boldsymbol{B}[n_2, n_2] \otimes F + \mathbb{I}_{n_2} \otimes Y_{GFM})^{-1}(\boldsymbol{B}[n_2, n_1] \otimes F) \quad (13)$$

$$\boldsymbol{Y}_N^{sub2} = (\boldsymbol{B}/n_1) \otimes F \quad (14)$$

where $\boldsymbol{B}[n_1, n_1]$ denotes the submatrix of $\boldsymbol{B}$ with rows indexed by $n_1$ and columns indexed by $n_1$, and $\boldsymbol{B}[n_1, n_2]$, $\boldsymbol{B}[n_2, n_1]$ and $\boldsymbol{B}[n_2, n_2]$ are all the submatrices of $\boldsymbol{B}$ with the same notation; $\boldsymbol{B}/n_1 = \boldsymbol{B}[n_2, n_2] - \boldsymbol{B}[n_2, n_1]\boldsymbol{B}^{-1}[n_1, n_1]\boldsymbol{B}[n_1, n_2]$ denotes the Schur complement of $\boldsymbol{B}[n_1, n_1]$ in $\boldsymbol{B}$ [14].

**Remark 1:** Since the smallest singular value naturally reflects the singularity degree of a matrix [12], we can derive $\underline{\sigma}\{\mathbb{I}_{2n} + \boldsymbol{L}^{\Gamma_3}\} = \min\{\underline{\sigma}\{\mathbb{I}_{2n_1} + \boldsymbol{L}^{sub1}\}, \underline{\sigma}\{\mathbb{I}_{2n_2} + \boldsymbol{L}^{sub2}\}\}$ based on (11). Then, the sensitivity peak of system $\Gamma_3$ in (4) can be rewritten as

$$\begin{aligned}\kappa_P^{\Gamma_3} &= \max_\omega \{1/\underline{\sigma}\{\mathbb{I}_{2n} + \boldsymbol{L}^{\Gamma_3}\}\}\\ &= \max_\omega \{1/\underline{\sigma}\{\mathbb{I}_{2n_1} + \boldsymbol{L}^{sub1}\}, 1/\underline{\sigma}\{\mathbb{I}_{2n_2} + \boldsymbol{L}^{sub2}\}\}\\ &= \max_\omega \{\kappa_P^{sub1}, \kappa_P^{sub2}\}.\end{aligned} \quad (15)$$

Thus, the DRP of the hybrid system $\Gamma_3$ can be equivalent to that of the two decoupled homogeneous subsystems 1 or 2 in (11).

Fig. 3 further shows the equivalence of system $\Gamma_3$ in Remark 1. It can be seen from Fig. 3 that subsystems 1 and 2 of system $\Gamma_3$ can be formed as homogeneous GFL and GFM inverters interconnected to the network admittance $\boldsymbol{Y}_N^{sub1}$ and $\boldsymbol{Y}_N^{sub2}$ instead of $\boldsymbol{Y}_N$, respectively. This inspired us to compare the DRP of system $\Gamma_3$ with that of systems $\Gamma_1$ and $\Gamma_2$ from a network-based perspective.

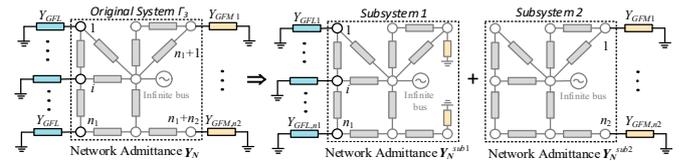

Fig. 3 Illustration of decoupled subsystems of system $\Gamma_3$ with the combination of GFL and GFM inverters.

The network admittance transfer function matrix of subsystem 1 in (13) can be further rewritten as [12]

$$\boldsymbol{Y}_N^{sub1} = (\boldsymbol{B}_{mod}/n_2) \otimes F \quad (16)$$

where $\boldsymbol{B}_{mod} = \boldsymbol{B} + 0 \oplus B_{eq}\mathbb{I}_{n_2}$ represents the modified susceptance matrix including the equivalent susceptance of GFM inverters; $B_{eq} > 0$ represents the equivalent susceptance of GFM inverters to approximate $Y_{GFM}$ in (13) at the interested frequency points [12]; $\oplus$ denotes the direct sum.

**Lemma 1:** The network grounded Laplacian matrix of subsystem 1 can be considered as $\boldsymbol{B}_{mod}/n_2$ in (16). The smallest eigenvalues of the network grounded Laplacian matrices of subsystem 1 and system $\Gamma_1$ satisfy

$$\underline{\lambda}\{\boldsymbol{B}_{mod}/n_2\} > \underline{\lambda}\{\boldsymbol{B}\} \quad (17)$$

*Proof.* The detailed proof is given in the Appendix B.

**Lemma 2:** The network grounded Laplacian matrix of subsystem 2 can be considered as $\boldsymbol{B}/n_1$ in (14). The largest eigenvalues of the network grounded Laplacian matrices in subsystem 2 and system $\Gamma_2$ satisfy

$$\bar{\lambda}\{\boldsymbol{B}/n_1\} < \bar{\lambda}\{\boldsymbol{B}\} \quad (18)$$

*Proof.* The detailed proof is given in the Appendix C.

**Remark 2:** For homogeneous GFL inverters, the DRP of system $\Gamma_1$ and subsystem 1 both can be quantified by the smallest eigenvalues of the network grounded Laplacian matrices, accordingly. Increasing smallest eigenvalue can be regarded as matching an increased SCR in the equivalent SIBS of the homogeneous GFL-based systems and improving its DRP. Based on Lemma 1, the smallest eigenvalue of

subsystem1 (i.e., $\underline{\lambda}\{\boldsymbol{B}_{\mathrm{mod}}/n_2\}$) is larger than that of system $\Gamma_1$ (i.e., $\underline{\lambda}\{\boldsymbol{B}\}$). Thus, the DRP of subsystem 1 is better than that of system $\Gamma_1$, that is,

$$\kappa_P^{sub1} = \max_\omega\{1/\underline{\sigma}\{\mathbb{I}_{2n_1}+\boldsymbol{L}^{sub1}\}\} < \kappa_P^{\Gamma_1}. \quad (19)$$

**Remark 3:** For homogeneous GFM inverters, the DRP of system $\Gamma_2$ and subsystem 2 both can be quantified by the largest eigenvalues of the network grounded Laplacian matrices, accordingly. Decreasing largest eigenvalue can be regarded as matching a decreased SCR in the equivalent SIBS of the homogeneous GFM-based systems and also improving its DRP. Based on Lemma 2, the largest eigenvalue of subsystem 2 (i.e., $\bar{\lambda}\{\boldsymbol{B}/n_1\}$) is smaller than that of system $\Gamma_2$ (i.e., $\bar{\lambda}\{\boldsymbol{B}\}$). Thus, the DRP of subsystem 2 is better than that of system $\Gamma_2$, that is,

$$\kappa_P^{sub2} = \max_\omega\{1/\underline{\sigma}\{\mathbb{I}_{2n_2}+\boldsymbol{L}^{sub2}\}\} < \kappa_P^{\Gamma_2}. \quad (20)$$

By combing Remarks 1~3, we illustrate that the synthesis of GFL and GFM inverters can more effectively ensure the DRP of multiple inverters for the general networks. In other words, the DRP of the hybrid system $\Gamma_3$ is proved to be at least superior to that of homogeneous systems $\Gamma_1$ and $\Gamma_2$ composed solely of either GFL or GFM inverters, as follows

$$\kappa_P^{\Gamma_3} = \max_\omega\{\kappa_P^{sub1},\kappa_P^{sub2}\} < \max_\omega\{\kappa_P^{\Gamma_1},\kappa_P^{\Gamma_2}\}. \quad (21)$$

## IV. CASE STUDY

Without the loss of generality, we take a three-inverter power system as the experimental system, as shown in Fig. 4. In such a system, three GFL/GFM inverters are connected at node 1~3; node 7 is specified as the infinite bus and node 4~6 are interior nodes of the grid. Parameters of inverters and networks are given in TABLE A.I in the Appendix A.

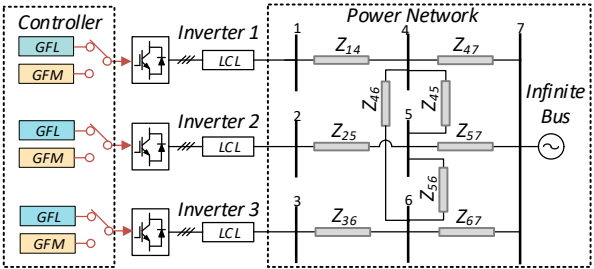

Fig. 4 One-line diagram of the three-inverter test system.

To validate the theoretical findings in Section III, we firstly construct systems $\Gamma_1$~$\Gamma_3$ by applying three combinations of GFL and GFM controllers to Inverter 1~3 as shown in TABLE. I ($n=3$, $n_1=1$, $n_2=2$). Then, two scenarios are created in the grid by setting a scaling parameter $k$ as 1.0 and 0.1, which is proportional to each line length in the per-unit system. The parameter is employed to uniformly decrease (or increase) all line impedances in the power network simultaneously. The frequency-domain curves of $\bar{\sigma}\{\boldsymbol{S}(j\omega)\}$ of systems $\Gamma_1$~$\Gamma_3$ in Scenario 1~2 are depicted in Fig. 5(a)~(b), respectively; the sensitivity peaks $\kappa_P$ as the DRP indicator are also listed in TABLE. I, accordingly.

The results in TABLE. I and Fig. 5 show that the sensitivity peaks of system $\Gamma_1$~$\Gamma_3$ in Scenario 1($k=1$) and Scenario 2($k=0.1$) both satisfy $\kappa_P^{\Gamma_3} < \max_\omega\{\kappa_P^{\Gamma_1},\kappa_P^{\Gamma_2}\}$. It is suggested that the synthesis of GFL and GFM inverters contributes to the low sensitivity peak of systems both in the weak and strong grids, thereby can more effectively ensure the DRP of multiple inverters. The observations from TABLE. I and Fig. 5 are consistent with those obtained from (21), which verifies the correctness of our analysis results in Section III.

The analysis results are further validated using hardware-in-the-loop (HIL) real-time simulations. The HIL setup includes a simulator, a digital controller (NI PXIe-1071), a host computer and an oscilloscope, as shown in Fig. 6. The HIL simulator, equipped with CPU and FPGA resources, enables high-fidelity simulations with a small step size (less than 0.5 μs). Here the grid and inverters (excluding controllers) are simulated using the FPGA. The controller for Inverter 1 is implemented in the digital controller, while the controllers for Inverter 2~3 are implemented on separate CPU cores in the HIL simulator.

TABLE. I
DRP QUANTIFICATION OF THREE INVERTER-BASED SYSTEMS IN TWO SCENARIOS

| System Cases | System $\Gamma_1$ | System $\Gamma_2$ | System $\Gamma_3$ |
|---|---|---|---|
| | GFLs at node 1-3 | GFMs at node 1-3 | GFLs at node 2 GFMs at node 1,3 |
| Scenario 1 ($k=1$) | | | |
| DRP Indicator | $\kappa_P^{\Gamma_1}=25.2$dB | $\kappa_P^{\Gamma_2}=10.8$dB | $\kappa_P^{\Gamma_3}=9.6$dB $<\kappa_P^{\Gamma_1}$ |
| Scenario 2 ($k=0.1$) | | | |
| DRP Indicator | $\kappa_P^{\Gamma_1}=3.2$dB | $\kappa_P^{\Gamma_2}=16.2$dB | $\kappa_P^{\Gamma_3}=13.7$dB $<\kappa_P^{\Gamma_2}$ |

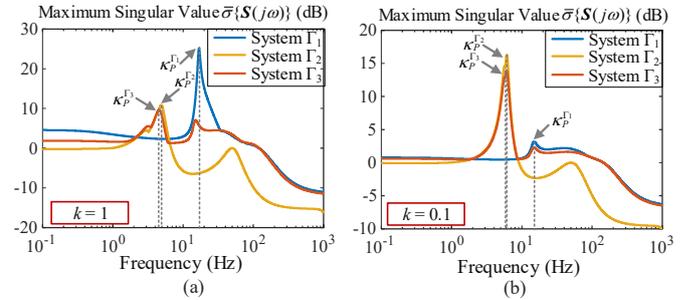

Fig. 5 Maximum singular values of sensitivity functions $\bar{\sigma}\{\boldsymbol{S}(j\omega)\}$ of system $\Gamma_1$~$\Gamma_3$ (a) in Scenario 1($k=1$); (b) in Scenario 2($k=0.1$).

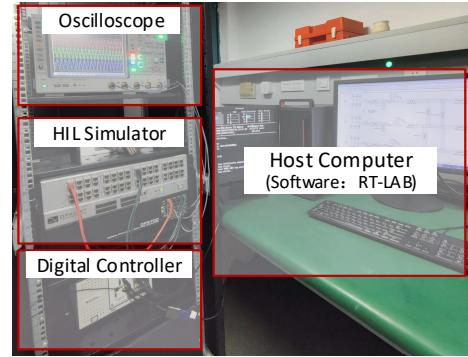

Fig. 6 HIL real-time simulation setup.

Fig. 7 (a) and (b) display the time-domain responses of systems $\Gamma_1$ and $\Gamma_3$ in Scenario 1($k=1$), respectively; Fig. 7 (c) and (d) display the time-domain responses of systems $\Gamma_2$ and $\Gamma_3$ in Scenario 2($k=0.1$), respectively. For systems $\Gamma_1$~$\Gamma_3$ in the two scenarios, the same 10% voltage dips as the external disturbance are applied to node 7 and then are cleared after 0.05s. It can be seen from Fig. 7 (a)-(d) that the system $\Gamma_3$ (with the synthesis of GFL and GFM inverters) is small-disturbance stable both in the two scenarios and has the better DRP than system $\Gamma_1$ or system $\Gamma_2$, which again validates the effectiveness of our results in Section III.

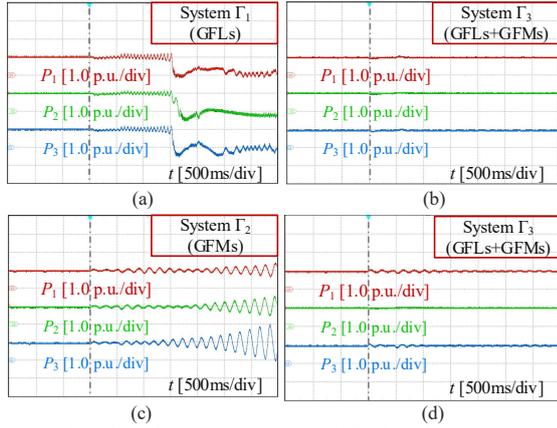

Fig. 7. Time-domain active power responses of the three-inverter systems under the same voltage disturbance. (a) and (b) are in Scenario 1($k$=1); (c) and (d) are in Scenario 2($k$=0.1).

## V. CONCLUSION

The paper analyzed the advantage of synthesizing GFL and GFM inverters to ensure the DRP of multi-inverter systems. It was demonstrated that the synthesis of GFL and GFM inverters could contribute to the low sensitivity peak of the systems, both in weak and strong grids. Hence, the DRP of multiple inverters could always be more effectively ensured than in homogeneous inverter-based systems that only use either GFL or GFM inverters. The results suggested that the necessity of implementing diverse inverters in power systems and potentially offered a basic guidance for planning and operating the large-scale integration of GFL and GFM inverters in the future grid.

## APPENDIX

### A. System Parameters

TABLE A.I
PARAMETERS OF INVERTERS AND POWER NETWORKS

| Base Values for per-unit Calculation |
|---|
| Voltage base value: $U_{base}$=0.69 kV   Power base value: $S_{base}$= 1.5MVA |
| Frequency base value: $f_{base}$ = 50Hz |
| Parameters of the Filter Part (per-unit values) |
| Inverter-side inductor: $L_f$= 0.05    LCL capacitor: $C_f$= 0.05 |
| Grid-side inductor: $L_g$= 0.05    R/L ratio of grid impedance: $\tau$ = 0.1 |
| Parameters of the GFL Controller (per-unit values) |
| PI parameters of the current control loop: 0.3, 10 |
| Voltage feedforward control: $K_{VF}$ = 1, $T_{VF}$= 0.004s |
| PI parameters of the active-reactive power control loop: 0.4, 8 |
| PI parameters of the phase lock loop (PLL): 20, 8020 |
| Parameters of the GFM Controller (per-unit values) |
| PI parameters of the current control loop: 0.3, 10 |
| Voltage feedforward control: $K_{VF}$ = 1, $T_{VF}$= 0.004s |
| PI parameters of the voltage control loop: 6, 20 |
| Parameters of virtual synchronous generator (VSG): $J$ = 2, $D$= 25 |
| Networks Parameters |
| $Z_{14}$ = 0.04+$j$0.05, $Z_{25}$ = 0.04+$j$0.05, $Z_{36}$ = 0.04+$j$0.05, |
| $Z_{45}$ = 0.02+$j$0.39, $Z_{46}$ = 0.02+$j$0.46, $Z_{56}$ = 0.02+$j$0.53, |
| $Z_{47}$ = 0.02+$j$0.53, $Z_{57}$ = 0.02+$j$0.19, $Z_{67}$ = 0.02+$j$0.46. |

### B. Proof of Lemma 1

Due to $B_{eq} > 0$ in (16), $\boldsymbol{B}_{mod}$ and $\boldsymbol{B}$ are positive-definite matrices and they satisfy $\boldsymbol{B}_{mod} = \boldsymbol{B} + 0 \oplus B_{eq}\mathbb{I}_{n2} > \boldsymbol{B}$. Then, the smallest eigenvalues of $\boldsymbol{B}_{mod}$ and $\boldsymbol{B}$ have $\underline{\lambda}\{\boldsymbol{B}_{mod}\} > \underline{\lambda}\{\boldsymbol{B}\}$. According to Theorem 2.1 in [14], the smallest eigenvalue of $\boldsymbol{B}_{mod}$ further has $\underline{\lambda}\{\boldsymbol{B}_{mod}/n_2\} > \underline{\lambda}\{\boldsymbol{B}_{mod}\} > \underline{\lambda}\{\boldsymbol{B}\}$, and thus conclude the proof ∎.

### C. Proof of Lemma 2

Refer to Theorem 2.1 in [14] and the largest eigenvalue of $\boldsymbol{B}$ has $\bar{\lambda}\{\boldsymbol{B}/n_1\} < \bar{\lambda}\{\boldsymbol{B}\}$. This concludes the proof ∎.

## REFERENCES

[1] Y. Li, Y. Gu, and T. C. Green, "Revisiting grid-forming and grid-following inverters: a duality theory," *IEEE Trans. Power Syst.*, vol. 37, no. 6, pp. 4541–4554, Nov. 2022.
[2] G. Weiss, Q.-C. Zhong, T. C. Green, and J. Liang, "H∞ repetitive control of DC-AC converters in microgrids," *IEEE Trans. Power Electron.*, vol. 19, no. 1, pp. 219–230, Jan. 2004.
[3] S. Skogestad and I. Postlethwaite, *Multivariable Feedback Control: Analysis and Design*. Hoboken, NJ, USA: Wiley, 2001.
[4] N. Hatziargyriou et al., "Definition and classification of power system stability-revisited & extended," *IEEE Trans. Power Syst.*, vol. 36, no. 4, pp. 3271–3281, Jul. 2021.
[5] IEEE Standard for interconnection and interoperability of inverter-based resources (IBRs) interconnecting with associated transmission electric power systems, IEEE Standard 2800TM-2022, 2022.
[6] F. Zhao, X. Wang and T. Zhu, "Power dynamic decoupling control of grid-forming converter in stiff grid," *IEEE Trans. Power Electron.*, vol. 37, no. 8, pp. 9073–9088, Aug. 2022.
[7] W. Dong, H. Xin, D. Wu, et al., "Small signal stability analysis of multi-infeed power electronic systems based on grid strength assessment," *IEEE Trans. Power Syst.*, vol. 34, no. 2, pp. 1393–1403, Mar. 2019.
[8] F. Ahmadloo and S. Pirooz Azad, "Grid interaction of multi-VSC systems for renewable energy integration," *IET Renewable Power Gen.*, vol. 17, no. 5, pp. 1212–1223, Apr. 2023.
[9] H. Pishbahar, F. Blaabjerg, and H. Saboori, "Emerging grid-forming power converters for renewable energy and storage resources integration – A review," *Sustainable Energy Technologies and Assessments*, vol. 60, p. 103538, Dec. 2023.
[10] Z. Zou, J. Tang, X. Wang, et al., "Modeling and control of a two-bus system with grid-forming and grid-following converters," *IEEE J. Emerg. Sel. Topics Power Electron.*, vol. 10, no. 6, pp. 7133–7149, Dec. 2022.
[11] F. Dorfler and F. Bullo, "Kron reduction of graphs with applications to electrical networks," *IEEE Trans. Circuits and Systems I: Regular Papers*, vol. 60, no. 1, Art. no. 1, Jan. 2013.
[12] H. Xin, Y. Wang, P. Ju, et al., "How many grid-forming converters do we need? a perspective from power grid strength," 2022, *arXiv*: 2209.10465.
[13] L. Huang et al., "Gain and phase: decentralized stability conditions for power electronics-dominated power systems," 2023, *arXiv*: 2309.08037.
[14] F. Zhang, *The Schur Complement and its Applications. in Numerical Methods and Algorithms*, no. 4. New York: Springer Science and Business Media, 2005.

**Fuyilong Ma** received the B.Eng. degree in electrical engineering from the College of Electrical Engineering, Zhejiang University, Hangzhou, China, in 2019. He is currently working toward the Ph.D. degree in the College of Electrical Engineering, Zhejiang University, Hangzhou. His research interests include renewable energy stability analysis and control.

**Huanhai Xin** (M'14) received the Ph.D. degree in College of Electrical Engineering, Zhejiang University, Hangzhou, China, in June 2007. He was a post-doctor in the Electrical Engineering and Computer Science Department of the University of Central Florida, Orlando, from June 2009 to July 2010. He is currently a Professor in the Department of Electrical Engineering, Zhejiang University. His research interests include distributed control in active distribution grid and micro-gird, AC/DC power system transient stability analysis and control, and grid-integration of large-scale renewable energy to weak grid.

**Zhiyi Li** (GSM'14-M'17) received the Ph.D. degree in Electrical Engineering from Illinois Institute of Technology in 2017. He received M.E. degree in Electrical Engineering from Zhejiang University (Hangzhou, China) in 2014 and the B.E. degree in Electrical Engineering from Xi'an Jiaotong University (Xi'an, China) in 2011. From August 2017 to May 2019, he was a senior research associate at Robert W. Galvin Center for Electricity Innovation at Illinois Institute of Technology. Since June 2019, he has been with the College of Electrical Engineering, Zhejiang University (Hangzhou, China) as a research professor. His research interests lie in the application of state-of-the-art optimization and control techniques in smart grid design, operation and management with a focus on cyber-physical security.

**Linbin Huang** (Member, IEEE) received the B.Eng. and Ph.D. degrees from Zhejiang University, Hangzhou, China, in 2015 and 2020, respectively. He is currently a Post-Doctoral Researcher with the Automatic Control Laboratory, ETH Zürich, Zürich, Switzerland. His research interests include power system stability, optimal control of power electronics, and data-driven control